\documentclass[preprint,article,accept,oneauthor]{Definitions/mdpi}
\firstpage{1} 
\pubvolume{1}
\issuenum{1}
\articlenumber{0}
\pubyear{2026}
\copyrightyear{2026}
\datereceived{ } 
\daterevised{ } 
\dateaccepted{ } 
\datepublished{ }

\usepackage{graphics,pstricks,pst-node}
\Title{An improved upper bound measure of star complexity of graphs}
\Author{Russell K. Standish}
\address{High Performance Coders, Sydney, Australia}
\corres{Correspondence: hpcoder@hpcoders.com.au}

\renewcommand{\star}{\mathrm{Star}}
\newcommand{\starbar}{\overline{\mathrm{Star}}}

\abstract{
  In \cite{Standish25c}, I explored the connection between star
  complexity and information based complexity. Because of the
  numerical difficulty in computing star complexity, I introduced a
  proxy measure that is an upper bound to star complexity, and showed
  a strong albeit non-linear relationship between the measures.
  In this paper, I introduce a tighter upper bound, by exploiting the
  well-known ABC package used to optimise logic circuits. In testing
  the new measure, I found that I had been computing the {\em formula
    complexity} variant of star complexity, rather than the tighter
  {\em circuit complexity} variant. Since Jukna clearly states the
  connection between star complexity and circuit complexity, I have
  modified the graph walking algorithm to capture circuit complexity
  rather than formula complexity.
  With this new ABC-based measure, applied to a set of 1000 500 vertex
  Erd\"os-Renyi graphs, a more linear relationship between star
  complexity and information based complexity is found.
}


\begin{document}
\nolinenumbers

\section{Introduction}

Information-based complexity (IBC) \cite{Standish01a} is a
well-defined complexity measure of any object given a description in a
language and a classifier that identifies those descriptions with the
object. When the descriptions are halting programs for a universal
Turing machine $U$, and the classifier is the output of $U$ when it
halts, the complexity measure is the negative logarithm of the
well-known Solomonoff-Levin distribution, and is closely related to
the notion of Kolmogorov complexity\cite{Li-Vitanyi97}.

Already in \cite{Standish01a} we see a generalisation of
Solomonoff-Levin complexity, where an observer of a system is
considered a classifier, and complexity of the system can be computed
by counting descriptions in some given language that the observer
considers to describe the system. Even though humans might be
considered to be Turing complete (Turing himself based his notion of
computability on what a human computer could do), in practical
circumstances, humans, and indeed any animal, will rely on short-cuts
to come to a conclusion about what any description means. The
risk of ``hanging'' whilst computing a non-halting description is
counter-productive to the business of living, and so the behaviour of
fully evaluating any description as a Turing machine would do will be
eliminated by evolution.

However, we hypothesise that given a particular classifier, that two
different encoding schemes will asymptotically give the same
complexity values for a given object, even if the classifiers aren't
acting as universal machines, provided the encoding schemes are
universal (able to generate descriptions for all entities in the domain
of interest).

In the current study, we have a graph IBC measure ${\cal C}$ developed
by encoding the edgelist, with the classifier being graph
automorphism\cite{Standish05a}. It turns out this measure is a specific
instance of one introduced by Mowshowitz\cite{Mowshowitz68c,Mowshowitz68d}.

One could compare this directly with star complexity, which is defined
as the minimal number of union and intersection operations of the
elementary stars required to construct the graph in question. In this
we need to draw the distinction between {\em formula complexity} and
{\em circuit complexity}, which is perhaps best illustrated by means
of an example:

\begin{pspicture}(-.5,-.5)(2.5,2.5)
  \rput(0,2){\rnode{1}{1}}
  \rput(2,2){\rnode{2}{2}}
  \rput(2,0){\rnode{3}{3}}
  \rput(0,0){\rnode{4}{4}}
  \rput(1,1){\rnode{0}{0}}
  \ncline01\ncline02\ncline03\ncline04
  \ncline12\ncline23\ncline34\ncline41
\end{pspicture}

\vspace{-2.5cm}
\hangindent=3.5cm
\hangafter=-11
The graph walker algorithm of \cite{Standish25c} finds the solution
\begin{displaymath}
  {\cal G}=(S_0\cup S_1\cup S_2)\cap((S_1\cup S_2)\cap S_0 \cup S_3 \cup S_4),
\end{displaymath}
which has an overall operation count of 7, where $S_i$ is the star
rooted at vertex $i$.

However, ABC finds the following formula:
\begin{displaymath}
  {\cal G}=S_0\cap (S_1\cup S_2\cup S_3\cup S_4)\cup(S_1\cup S_2)\cap(S_3\cup S_4), 
\end{displaymath}
which whilst looking like it uses 8 operations (``formula complexity'')
can reuse the common calculation of $S_1\cup S_2$ and $S_3\cup S_4$ to
give an algorithm using only 6 operations:
\begin{eqnarray*}
  x_1&=&S_1\cup S_2\\
  x_2&=&S_3\cup S_4\\
  {\cal G}&=&S_0\cap (x_1\cup x_2)\cup x_1\cap x_2
\end{eqnarray*}
Because this is not a single formula, this is known as the {\em
  circuit complexity}. Jukna\cite{Jukna13} notes the connection of star
complexity with circuit complexity.

\section{ABC optimised star complexity upper bound,
  $\starbar_{ABC}({\cal G})$}

ABC is a package for optimisation and validation of logic
circuits\cite{Brayton-Mischenko10,Fan-Wu23}. To use it, we used the C
API to create a single-valued logic function with $n$ boolean inputs,
where $n$ is the number of vertices. Initially, this was seeded with the
or of all stars covered by the graph, and the remaining edges in the
form $S_i\cap S_j$ for and edge joining vertex $i$ to vertex $j$. Then the
resyn3\cite{Li-etal24} sequence of operations was applied until the
gate count was no longer reduced, or up to a fixed maximum number of
times (100), to prevent the algorithm from having exponential complexity.

However, it was found that it didn't always find a solution better
than the $\starbar$ algorithm outlined in \cite{Standish25c}, so it
was modified to be seeded by the output of that algorithm, with one
minor modification. The algorithm as described in \cite{Standish25c}
chose one of the largest vertex degree vertices to start the collection of
terms into factors, which may not be the optimal vertex. The algorithm
was modified to to evaluate each of the maximal degree vertices in turn,
and choose the best. The process was not repeated at the next
iteration, to prevent combinatorial explosion.

\section{Walking the space of graphs}

The code used for walking formulae for graphs used a reverse polish
notation (RPN) to represent the formula, with symbols corresponding to
pushing a star onto the stack, and symbols corresponding to taking the
bitwise and or or of the top two elements of the stack, and placing
the result on the top of the stack. To enable reuse of previously
computed values, I added two more symbols, corresponding to
duplicating the top element of the stack and to rotating the stack so
that the top element moves to the bottom.

For example, the new RPN code can generate the previously mentioned
example graph as
\begin{displaymath}
  01\cup\Downarrow2\cup3\Downarrow\Updownarrow\cap\Updownarrow\cup4\cap\cup
\end{displaymath}
where $\Downarrow$ means duplicate the top element of the stack, and
$\Updownarrow$ means circularly shifting the stack up by one element,
leaving the top element at the bottom. As can be seen, the RPN formula
has the required 6 $\cup$ and $\cap$ operations.

For walking formulae corresponding to graphs with a particular star
value $s$, we walk the space of strings
$\{0,\ldots,n-1,\cap,\cup,\Downarrow,\Updownarrow\}^\ell$, where the
length $\ell$ depends on the star value. Since there must be
$\mathrm{star}$ $\cap$ and $\cup$ operations, which decrease the stack
and the symbols $\{0,\ldots,n-1,\Downarrow\}$ which increase the
stack, there must be exactly $s$ symbols that decrease the stack, and
$s+1$ symbols that increase it. Finally $\Updownarrow$ neither
increases nor decreases the stack, but it makes sense to only rotate
the stack after a logic operation, or duplication operation, so we
iterate of all $2^{2s-1}$ possible combinations of rotating the stack,
  or not after each logical operation.

We can also exploit the fact that vertex labels are arbitrary, so the
first star pushed to the stack is always 0, the second always 1, then
after that from the range $\{0,1,2\}$ etc, which reduces the number of
strings to examine.

The code implementing this new algorithm, plus $\starbar_{ABC}$ can be
found as release 1.0 of the
starcomplexity\footnote{https://github.com/highperformancecoder/starcomplexity}
project on Github. Data from the runs is available in suplementary
data, available from the Open Science Framework\cite{Standish26a}.
  
\begin{table}
  \begin{center}
  \begin{tabular}{|rr|rrrrrrrrrr|}
      \hline
      \hspace*{1.5ex}&&&&&&&\rput(0,0){$\starbar_{ABC}$}&&&&\\
      &&0&1&2&3&4&5&&7&8&9\\\hline
      &0&1&&&&&&&&&\\
      &1&&3&&&&&&&&\\
      &2&&&4&&&&&&&\\
      &3&&&&6&&&&&&\\
      \rput{-90}(0,0){Star}
      &4&&&&&12&&&&&\\
      &5&&&&&&25&&&&\\
      &6&&&&&&&45&3&&\\
      &7&&&&&&&&96&17&1\\\hline
  \end{tabular}
  \end{center}
  \caption{Counts of graphs with particular Star and $\starbar_{ABC}$
    values, out of all 10 vertex graphs with $\star\le 7$. Most graphs
  lie along the $\star=\starbar_{ABC}$ diagonal.}
  \label{star-starbarABC-counts}
\end{table}

\begin{table}
  \begin{center}
    \begin{tabular}{|rr|rrrrrrrrrrrr|}
      \hline
      \hspace*{1.8ex}&&&&&&&\rput(0,0){$\starbar$}&&&&&&\\
      &&0&1&2&3&4&5&6&7&8&9&10&11\\\hline
      &0&1&&&&&&&&&&&\\
      &1&&3&&&&&&&&&&\\
      &2&&&4&&&&&&&&&\\
      &3&&&&6&&&&&&&&\\
      \rput{-90}(-.3em,0){$\starbar_{ABC}$}
      &4&&&&&12&&&&&&&\\
      &5&&&&&&22&3&&&&&\\
      &6&&&&&&&37&5&3&&&\\
      &7&&&&&&&&72&16&8&3&\\
      &8&&&&&&&&1&6&6&4&\\
      &9&&&&&&&&&&&&1\\\hline
    \end{tabular}
  \end{center}
    \caption{Counts of graphs with particular $\starbar_{ABC}$ and $\starbar$
    values, out of all 10 vertex graphs with $\star\le 7$. It
    illustrates that $\starbar_{ABC}\le\starbar$ mostly (with one
    exception in this table)}
  \label{starbar-starbarABC-counts}
\end{table}

\section{$\starbar_{ABC}$ versus ${\cal C}$ of randomly generated
  Erd\"os-Renyi graphs}

We repeat the experiment outlined in \cite{Standish25c}, generating
1000 500-vertex Erd\"os-Renyi graphs (instead of the 1000-vertex graphs
generated in \cite{Standish25c}), and computing ${\cal C}$, the
improved $\starbar$ and $\starbar_{ABC}$. The results can be seen in
figure {ER-500}.

\begin{figure}
  \psset{xunit=12mm,yunit=6mm}
  \begin{pspicture}(10,10)
    \rput(5,5.5){\resizebox{9.5\psxunit}{9.5\psyunit}{\includegraphics{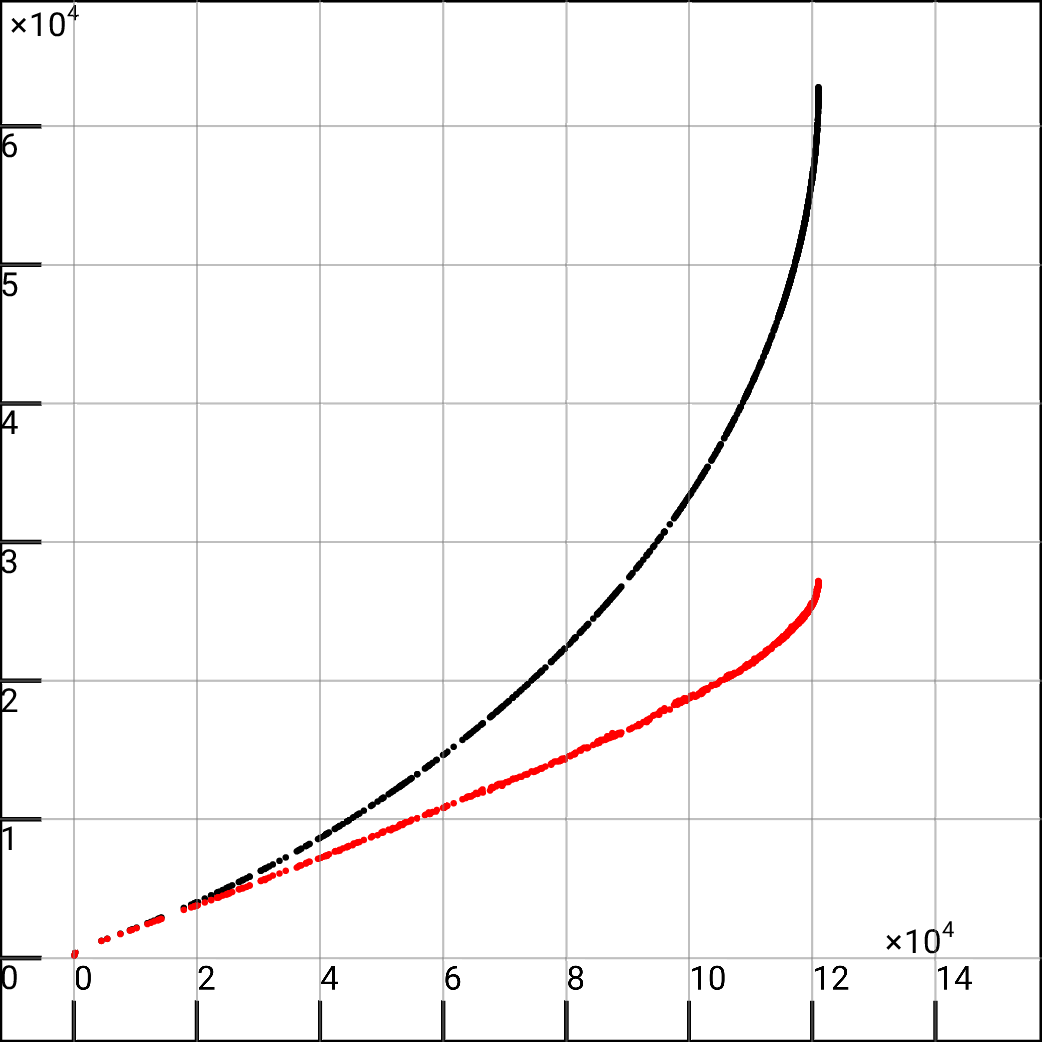}}}
    \rput(2.5,8){$\starbar$}
    \rput(2.5,7){{\red $\starbar_{ABC}$}}
    \psline{->}(3,8)(7,8)
    \psline[linecolor=red]{->}(3,7)(6,4)
    \rput(5,0){${\cal C}$}
  \end{pspicture}
  \caption{Plot of $\starbar$ and $\starbar_{ABC}$ against ${\cal C}$ for 1000 Erd\"os-Renyi
  randomly generated graphs of 500 vertices each.}
  \label{ER-500}
\end{figure}

As expected, $\starbar_{ABC}$ is a much tighter upper bound than the
original algorithm, but also more interestingly, it has a far more
linear relationship against information-based complexity $C$. The
departure from linearity occurs for the most complex graphs, and one
may speculate that ABC has most difficulty in optimising the circuit
complexity of those more complex graphs. 

\section{Conclusion}

In this paper, we explore estimating star complexity of graphs by
means of the ABC package, widely used for optimising electronic logic
circuits. It proves to be a more accurate estimate of star complexity
than the proposal in \cite{Standish25c}, and also demonstrates an even
stronger relationship between star complexity and information based complexity.

\bibliography{rus}
\end{document}